\documentclass{ws-ijmpd}
\usepackage[super,compress]{cite}
\begin{document}

\markboth{A.~Oliveros, Marcos A.~Jaraba}
{{Inflation driven by massive vector fields}}

%
\catchline{}{}{}{}{}
%

\title{Inflation driven by massive vector fields\\
with derivative self-interactions}

\author{\footnotesize A.~OLIVEROS$^{*}$,  MARCOS A.~JARABA$^{\dagger}$}

\address{Programa de F\'isica, Universidad del Atl\'antico, \\ Carrera 30 N\'umero 8-49 Puerto Colombia-Atl\'antico, Colombia\\
$^{*}$alexanderoliveros@mail.uniatlantico.edu.co\\
$^{\dagger}$markjarava@gmail.com}

\maketitle


\begin{abstract}
Inspired by the Generalized Proca Theory,  we study a  vector-tensor model of inflation with  massive vector fields and derivative self-interactions. The action under consideration contains a usual Maxwell-like kinetic term, a general potential term and a term  with non-minimal derivative coupling between the vector field and gravity, via the dual Riemann tensor. In this theory, the last term contains  a free parameter, $\lambda$, which quantifies the non-minimal derivative coupling. In this scenario, taking into account a spatially flat FRW  universe and a general  vector field, we obtain the general expressions for the equation of motion and the  total energy momentum tensor.  Choosing a  Proca-type potential, a suitable inflationary regimen driven by massive vector fields is studied. In this model, the isotropy of expansion  is guaranteed by considering a triplet of orthogonal vector fields. In order to obtain an inflationary solution with this model, the quasi de Sitter expansion was considered. In this case the vector field  behaves as a constant. Finally,  slow roll analysis is performed and slow-roll conditions are defined for this model, which, for suitable constraints of the model parameters, can give the required number of e-folds for sufficient inflation.
\end{abstract}


\ccode{PACS numbers: 98.80.Cq, 98.80.-k, 04.50.Kd}


\section{Introduction}\label{sec_intro}

The hypothesis of inflation was introduced in cosmology in the early 80's to resolve some problems present in the Hot Big Bang model of the universe
(e.g., the horizon, the flatness, and the monopole problems, among others). \cite{guth, albrecht, linde} In the inflationary regimen, the space grows exponentially (or quasi-exponentially) fast in a very short time and at very high energy scales after the Big Bang. Furthermore, inflation  can explain other phenomena observed in the current universe: the temperature fluctuations in CMB spectrum \cite{hu, soda, wands}, the existence of large scale structures \cite{chibisov, hawking, guth2}, and the nearly scale invariant primordial power spectrum \cite{lyth, lidsey}.\\

\noindent Usually, to explain the dynamics of inflation it is necessary to introduce one or more scalar fields coupled to gravity. In these models, the fundamental condition requires that the potential energy of the fields dominate over their kinetic term, namely, the potential should be flat. Under this condition, it has the so-called slow-roll inflation (for more details about this topic see  Refs.~\citen{linde2}
and \citen{liddle1}). Besides the scalar fields models of inflation, in recent years some authors have considered another alternative, suggesting the possibility that the inflation is driven  by  massive vector fields \cite{golovnev, setare, darabi, koh, oliveros1}. The vector fields  also  have been considered as another alternative to solve the dark energy problem \cite{picon, kiselev, wei, mota, maroto1, maroto2, harko, oliveros2}. The most general theory for a massive vector field with derivative self-interactions with a priori arbitrary coefficients, was introduced in Refs.~\citen{lavinia1} and \citen{tasinato}, the so-called,  Generalized Proca Theory. This model constitutes the Galileon-type generalization of the Proca action with only three propagating physical degrees of freedom.The Generalized Proca Theory and its extensions (Beyond generalized Proca theories) have been extensively studied in the literature in the last years 
\cite{allys1,lavinia3, lavinia4, lavinia5, kimura, allys2, allys3, lavinia6, nakamura1, lavinia7, lavinia8, nakamura2, emami}. On the other hand, 
the gravitational-wave event GW170817 from a binary neutron star merger together with the electromagnetic counterpart showed that the speed of gravitational waves (GW) $c_t$ is very close to that of light for the redshift $z<0.009$ (see  Refs.~\citen{abbot1} and \citen{abbot2}). This places tight constraints on dark energy models (DE) constructed in the framework of modified gravitational theories (since these theories predict a variable GW speed at low redshift). From the above, in order to guarantee the viability (at low redshift) of the   Horndeski theories and its generalizations (e.g, the Generalized Proca Theory) a rigorous analysis must be performed   (for a recent review about this topic see Ref.~\citen{kase1}). Nevertheless, this restriction does not apply to high redshift values, thus, these modified gravitational theories could be used in an inflationary context.\\

\noindent Recently,  a new proposal has been introduced in Ref.~\refcite{lavinia9} (dubbed Scalar-Vector-Tensor Gravity Theories), where the authors constructs a consistent ghost-free covariant scalar-vector-tensor gravity theories with second order equations of motion with derivative interactions, and in this scenario, it is possible to unify Horndeski and generalized Proca theories.  The generalized
Proca interactions can be recovered taking the corresponding limit of the free general functions  accordingly. These scalar-vector-tensor theories could have important applications  in cosmology and astrophysics, namely: inflation, new black hole and neutron star solutions, dark matter and dark energy \cite{lavinia10, lavinia11, kase}.\\ 

\noindent In this work, taking into account the Generalized Proca Theory,  we consider a  vector-tensor model of inflation with  massive vector fields and derivative self-interactions. The action under consideration contains a usual Maxwell-like kinetic term, a general potential term and a term  with non-minimal derivative coupling between the vector field and gravity, via the dual Riemann tensor. In this theory, the last term contains  a free parameter, $\lambda$, which quantify the non-minimal derivative coupling.  The problem with the spatial anisotropy  is solved considering a triplet of orthogonal vector fields \cite{picon, golovnev}.\\

\noindent  This paper it is organized as follows: in section \ref{sec_model} we introduce the vector-tensor model of inflation with  massive vector fields and derivative self-interactions, and the  corresponding field equations are obtained. In section \ref{sec_cosmo}, we consider a flat FRW  universe and a vector field with only spatial components, and from these considerations  general expressions for equation of motion and the  total energy momentum tensor are obtained. Also, in this section we consider two possible inflationary scenarios driven by massive vector fields: a quasi de Sitter expansion and the slow-roll approximation.  Finally, some conclusions are exposed in section \ref{sec_concs}.

\section{The model}\label{sec_model} In this section we shall present in brief some basic features of the  generalized Proca theories, and also, we derive the general equations for the model under consideration.\\

\noindent  The most general vector-tensor model with second-order equations of motion (the so-called generalized Proca theories) is given by the action \cite{lavinia1, tasinato}
\begin{equation}\label{eq1}
S_{\text{gen.Proca}}=\int{d^4x\,\sqrt{-g}\sum_{i=2}^{6}{\mathcal{L}_i}},
\end{equation}
where $g$ is the determinant of the metric $g_{\mu\nu}$, and the terms $\mathcal{L}_i$ ($i=2,3,4,5,6$) are given by
\begin{equation}\label{eq2}
\mathcal{L}_2=G_2(A_{\mu}, F_{\mu\nu},\tilde{F}_{\mu\nu})=G_2[X,F^2,F\cdot \tilde{F},(A\cdot\tilde{F})^2],
\end{equation}
\begin{equation}\label{eq3}
\mathcal{L}_3=G_3(X)\nabla_{\mu}A^{\mu},
\end{equation}
\begin{equation}\label{eq4}
\mathcal{L}_4=G_4(X)R+G_{4,X}(X)[(\nabla_{\mu}A^{\mu})^2-\nabla_{\mu}A_{\nu}\nabla^{\nu}A^{\mu}],
\end{equation}
\begin{equation}\label{eq5}
\begin{aligned}
\mathcal{L}_5=&G_5(X)G_{\mu\nu}\nabla^{\mu}A^{\nu}-\frac{1}{6}G_{5,X}(X)[(\nabla_{\mu}A^{\mu})^3
-3\nabla_{\mu}A^{\mu}\nabla_{\rho}A_{\sigma}\nabla^{\sigma}A^{\rho}\\
&+2\nabla_{\rho}A_{\sigma}\nabla^{\nu}A^{\rho}\nabla^{\sigma}A_{\nu}]
-g_{5}(X)\tilde{F}^{\alpha\mu}\tilde{F}_{\mu}^{\beta}\nabla_{\alpha}A_{\beta},
\end{aligned}
\end{equation}
\begin{equation}\label{eq6}
\mathcal{L}_6=G_6(X)L^{\mu\nu\alpha\beta}\nabla_{\mu}A_{\nu}\nabla_{\alpha}A_{\beta}
+\frac{1}{2}G_{6,X}(X)\tilde{F}^{\alpha\beta}\tilde{F}^{\mu\nu}\nabla_{\alpha}A_{\mu}\nabla_{\beta}A_{\nu}.
\end{equation}
where $X=-\frac{1}{2}A_{\mu}A^{\mu}$, $F^2=F_{\mu\nu}F^{\mu\nu}$, $F\cdot \tilde{F}=F_{\mu\nu}\tilde{F}^{\mu\nu}$, $(A\cdot\tilde{F})^2=A_\alpha\tilde{F}^{\alpha\sigma}A_\beta\tilde{F}^{\beta}{\sigma}$, ($F_{\mu\nu}=\nabla_{\mu}A_{\nu}-\nabla_{\nu}A_{\mu}$),  $G_{i,X}=\partial G_i/\partial X$, $\tilde{F}^{\mu\nu}=\epsilon^{\mu\nu\alpha\beta}F_{\alpha\beta}/2$ is the dual strength tensor, $R$ is the Ricci scalar, $G_{\mu\nu}$ is the Einstein tensor and
$L^{\mu\nu\alpha\beta}$ is the double dual Riemann tensor, defined by $L^{\alpha\beta\gamma\delta}=-\frac{1}{2}\epsilon^{\alpha\beta
\mu\nu}\epsilon^{\gamma\delta\rho\sigma}R_{\mu\nu\rho\sigma}$, where \linebreak $\epsilon_{\alpha\beta\mu\nu}$ is the Levi-Civita tensor.
In Eq. (\ref{eq2}), $G_2$ is an arbitrary function of all possible scalars made out $A_\mu$, $F_{\mu\nu}$ and $\tilde{F}_{\mu\nu}$ of containing both parity violating and preserving terms \cite{allys3}.\\

\noindent  From the Generalized Proca Model given by (\ref{eq1}), it is possible to perform some choices of the  terms given by $\mathcal{L}_i$  with arbitrary or particular coupling functions $G_i$ ($i=2,3,4,5,6$). For example,  in  Ref.~\refcite{emami} the authors build an inflationary model whose action contains the terms $\mathcal{L}_2$ and $\mathcal{L}_4$ only. In this work, we consider for the action given by Eq.~(\ref{eq1}) the following special case
\begin{equation}\label{eq7}
S=S_{\text{H-E}}+\int{d^4x\,\sqrt{-g}(\mathcal{L}_2+\mathcal{L}_6)},
\end{equation} 
where $S_{\text{H-E}}$ stands the Hilbert-Einstein action. Besides, for simplicity we take $\mathcal{L}_2=G_2[X,F^2,F\cdot \tilde{F},(A\cdot\tilde{F})^2]=-\frac{1}{4}F^2+V(X)$ and $G_6(X)=\lambda$ (namely: $G_{6,X}(X)=0$). Thereby, the action given by the Eq.~(\ref{eq7}) 
takes the following explicit form
\begin{equation}\label{eq8}
\begin{aligned}
S=&\int d^4x\,\sqrt{-g}\Bigg(\frac{M_{\text{p}}^2}{2}R-\frac{1}{4}F_{\mu\nu}F^{\mu\nu}+V(X)
+\lambda L^{\mu\nu\alpha\beta}F_{\mu\nu}F_{\alpha\beta}\Bigg)\\
&=\int d^4x\,\sqrt{-g}\Bigg[\frac{M_{\text{p}}^2}{2}R-\frac{1}{4}F_{\mu \nu}F^{\mu \nu}+V(X)-\\
&\frac{\lambda}{4}\Big( R F_{\mu \nu}F^{\mu \nu}-
4 R_{\mu \nu} F^{\mu \sigma} F_\sigma^\nu+R_{\mu \nu \alpha \beta}F^{\mu \nu}F^{\alpha \beta} \Big)\Bigg],
\end{aligned}
\end{equation} 
where $M_{\text{p}}$ in the reduced Planck mass, $R$ is the Ricci scalar and $\lambda$ is a parameter with dimensions $M^{-2}$,
moreover, the explicit form for the double dual Riemann tensor has been introduced. 
A similar action to Eq.~(\ref{eq8}) (without potential term) have been studied in other contexts, for example in Ref.~\refcite{barrow}, the authors have focused their analysis  on  a dynamical setup which is well understood in the conventional electromagnetism  in an axisymmetric Bianchi type I universe.
Also, they have considered  the cosmological consequences of this action taking into account the observational constraints and
the sign of the parameter $\lambda$. In Ref.~\refcite{lavinia2}, the authors have explored the model  in a de Sitter spacetime
and studied the presence of instabilities and shown that it corresponds to an attractor solution in the presence of the vector field.
Furthermore,  they have investigated the cosmological evolution and stability of perturbations in a general FRW spacetime
(namely, absence of ghosts and Laplacian instabilities in this theory). Finally, in Ref.~\refcite{davydov} the authors propose a new mechanism for inflation using classical SU(2) Yang-Mills homogeneous and isotropic field non-minimally coupled to gravity via the dual Riemann tensor, however in Ref.~\refcite{lavinia12} the authors perform a perturbative analysis, showing that this model does not provide a
consistent inflationary framework due to the presence of ghosts and Laplacian instabilities.\\  

\noindent The variation of the action, Eq.~(\ref{eq8}), with respect to the metric tensor $g_{\mu\nu}$ gives the field equations
\begin{equation}\label{eq9}
R_{\mu \nu} - \frac{1}{2}g_{\mu \nu}R = \kappa^2 T_{\mu \nu},
\end{equation}
where $\kappa^2=8\pi G=M_p^{-2}$ and $T_{\mu \nu}$ is the energy momentum tensor for the vector field, which  we can split in
five terms (for pedagogical purposes), as follows:
\begin{equation}\label{eq10}
T_{\mu \nu} = T_{\mu \nu}^{(F^2)}+T_{\mu\nu}^{(V)}+T^{(1)}_{\mu \nu}+T^{(2)}_{\mu \nu}+T^{(3)}_{\mu\nu}.
\end{equation}
In particular, each term is given by the following expressions:
\begin{equation}\label{eq11}
T_{\mu \nu}^{(F^2)}=F_\mu^\sigma F_{\sigma \nu}-\frac{1}{4}g_{\mu \nu}F_{\sigma \rho}F^{\sigma \rho},
\end{equation}
\begin{equation}\label{eq12}
T_{\mu\nu}^{(V)}=-2 \frac{dV}{dX}A_\mu A_\nu+g_{\mu \nu}V(X),
\end{equation}
\begin{equation}\label{eq13}
T_{\mu \nu}^{(1)}=-\frac{ \lambda}{2} \Big[ \nabla_\mu \nabla_\nu\Big(F_{\alpha \beta}F^{\alpha \beta}\Big)-g_{\mu \nu}\square\Big(F_{\alpha \beta}F^{\alpha \beta}\Big)
+2 RF_\mu^{\sigma}F_{\nu \sigma }-2 F_{\alpha \beta}F^{\alpha \beta}G_{\mu \nu}\Big],
\end{equation}
\begin{equation}\label{eq14}
\begin{aligned}
T_{\mu \nu}^{(2)}=&\lambda \square\Big(F_\mu^\sigma F_{\nu \sigma}\Big)+g_{\mu \nu}\Big[\nabla_\rho\nabla_\gamma\Big(F^{ \sigma \rho }F_{\sigma}^\gamma\Big)-
R_{\alpha \beta}F^{ \sigma \alpha }F_{\sigma}^\beta\Big]+2\nabla_\rho\nabla_{(\mu}\Big(F_{\nu) \sigma }F^{\rho \sigma }\Big)-\\
&2R_{\gamma \rho}F_{\mu}^\gamma F_\nu^{\rho}+4F^{\gamma \sigma}F_{\sigma (\mu}R_{\nu) \gamma}, 
\end{aligned}
\end{equation}
\begin{equation}\label{eq15}
T_{\mu \nu}^{(3)}=-\frac{\lambda}{4}\Big[4\nabla^\gamma\nabla_{\sigma}\Big(F_{(\mu \gamma}F_{\nu)}^\sigma\Big)-g_{\mu \nu}R_{\rho \alpha \beta}^\gamma F_\gamma^{ \rho }F^{ \alpha \beta }-\\
6  F^{ \alpha \beta } F_{ \gamma (\mu}R_{\nu) \alpha \beta}^\gamma \Big].
\end{equation}
\noindent On the other hand, the variation of the action with respect to the vector field, gives the equation of  motion 
\begin{equation}\label{eq16}
\begin{aligned}
&-\Big(1+\lambda R \Big)\nabla_\mu F^{\mu \nu}
-\lambda F^{\mu \nu}\nabla_\mu R-4\lambda\nabla_\mu\Big( F^{\sigma [\mu}R_\sigma^{\nu]}\Big)+\\
&\lambda\nabla^\mu\Big(R_{\mu\sigma \rho}^\nu F^{\sigma \rho}\Big)-2V'(X) A^\nu=0,
\end{aligned}
\end{equation}
where the prime denotes the derivative respect to $X$ variable.\\

\section{Cosmological analysis}\label{sec_cosmo}
\noindent In this section, we study a possible inflationary regimen of the universe using the background equations obtained in the previous section. In this sense, the first step is to consider a flat, homogeneous and isotropic universe whose metric  is given by Friedmann-Robertson-Walker (FRW) metric:
\begin{equation}\label{eq17}
ds^2=-dt^2+a(t)^2 \delta_{ij}dx^idx^j,
\end{equation}
where $a(t)$ is the scale factor.  In addition, we consider that the vector field has the spatial components only, namely, $A_\mu=(0,A_i)$.
The contribution to the basic equations of the model from the temporal component is null. \\

\noindent The explicit form for the total energy-momentum tensor $T_{\mu\nu}$ is obtained adding the tensors given by the Eqs.~(\ref{eq11}), (\ref{eq12}), (\ref{eq13}), (\ref{eq14}),
(\ref{eq15}). Using the FRW metric, the component $T_0^0$ is:
\begin{equation}\label{eq18}
T_0^0=\frac{1}{4}F^2\left(1+6\lambda H^2\right)+V,
\end{equation}
and the spatial components $T_j^i$ are:
\begin{equation}\label{eq19}
\begin{aligned}
T_j^i=&\frac{1}{a^2}\Big[\lambda H \partial_t \Big(\dot{A}_i\dot{A}_j\Big)+\Big(\lambda \dot{H}-\lambda H^2-1-2V'(X)\Big)A_i A_j\Big] + \\
&\frac{1}{4}\Big[2\lambda H \dot{F}^2+\Big(2\lambda \dot{H}+4\lambda H^2-1\Big)F^2+V \Big]\delta_j^i,
\end{aligned}
\end{equation}
It is evident from Eq.~(\ref{eq19}) that the spatial part for the total energy-momentum tensor
is not diagonal, therefore, the spatial isotropy is broken.\\

\noindent In a similar way, the equation of motion for the vector field (\ref{eq16}) is given by  
\begin{equation}\label{eq20}
\left(1+2\lambda H^2\right)\ddot{A}_i+H\left(1+4\lambda \dot{H}+2\lambda H^2\right)\dot{A}_i-2V' (X)A_i=0.
\end{equation}

\noindent To restore the isotropy, it is considered a triplet of mutually orthogonal vector
fields ($\vec{A}^{(1)}, \vec{A}^{(2)},\vec{A}^{(3)}$) \cite{picon, golovnev}, whose components are given by:
\begin{equation} \label{eq21}
A_\mu^{(n)}=a(t)A(t)\delta_\mu^n,
\end{equation}
with $n=\{1,2,3\}$. $A(t)$ represents the magnitude of the vector fields.
The orthogonality of the vector fields is given by
\begin{equation}\label{eq22}
\vec{A}^{(n)}\cdot \vec{A}^{(m)}=A^2 \delta_m^n,
\end{equation}
additionally
\begin{equation}\label{eq23}
\sum_{n=1}^3{A_i^{(n)}A_j^{(n)}}=A^2 \delta_j^i.
\end{equation}
From these relations,  Eqs.~(\ref{eq18}) and (\ref{eq19}) take the following
form 
\begin{equation}\label{eq24}
T_0^0=-\frac{3}{2}\left(1+6\lambda H^2\right)\left(\dot{A}+AH\right)^2 +3V,
\end{equation}
and 
\begin{equation}\label{eq25}
\begin{aligned}
T_j^i=&\Big\{-\frac{1}{2}\Big(\dot{A}+AH\Big)\Big[8\lambda H \ddot{A}+\Big(18\lambda H^2+4\lambda \dot{H}-1\Big)\dot{A}+\\
&\Big(10\lambda H^3+12\lambda H \dot{H}-H\Big)A\Big]+\Big(3V-2V'(X) A^2\Big)\Big\} \delta_j^i,
\end{aligned}
\end{equation}
Similarly, the Friedmann equations take the following form:
\begin{equation}\label{eq26}
2H^2=\kappa^2\left[\left(1+6\lambda H^2\right)\left(\dot{A}+AH\right)^2-2V\right],
\end{equation}
and
\begin{equation}\label{eq27}
\begin{aligned}
&6H^2+4\dot{H}=\kappa^2\Big(\dot{A}+AH\Big)\Big[8\lambda H \ddot{A}+\Big(18\lambda H^2+4\lambda \dot{H}-1\Big)\dot{A}+\\
&\Big(10\lambda H^3+12\lambda H \dot{H}-H\Big)A\Big]+2\kappa^2\Big(2V'(X)A^2-3V\Big),
\end{aligned}
\end{equation}
finally, the equation of motion reads:
\begin{equation}\label{eq28}
\begin{aligned}
&\Big(1+2\lambda H^2\Big)\ddot{A}+\Big(3H+4\lambda H \dot{H}+6\lambda H^3\Big)\dot{A}+\\
&\Big(2H^2+4\lambda H^4+\dot{H}+6\lambda \dot{H}H^2-2V'(X)\Big)A=0.
\end{aligned}
\end{equation}
\subsection{Inflation driven by massive vector fields}
\noindent Let's now consider a particular form for the potential $V(X)$. In the literature, the usual choice  for the potential is
$V(X)=-\frac{1}{2}m^2A_{\mu}A^{\mu}$ (Proca potential), wich represents a massive vector field with mass $m$. In this case, the associated gauge invariance of the vector field is explicitly broken. However, it can be reintroduced in a straightforward manner using the Stueckelberg trick.\\

\noindent  With this choice for the potential, the Friedmann equations (\ref{eq26}) and (\ref{eq27}) are reduce to:
\begin{equation}\label{eq29}
2H^2=\kappa^2\left[\left(1+6\lambda H^2\right)\left(\dot{A}+AH\right)^2 +m^2A^2\right],
\end{equation}
and
\begin{equation}\label{eq30}
\begin{aligned}
6H^2+4\dot{H}=&\kappa^2\Big(\dot{A}+AH\Big)\Big[8\lambda H \ddot{A}+\Big(18\lambda H^2+4\lambda \dot{H}-1\Big)\dot{A}+\\
&\Big(10\lambda H^3+12\lambda H \dot{H}-H\Big)A\Big]+\kappa^2m^2 A^2,
\end{aligned}
\end{equation}
The equation of motion (\ref{eq28}) takes the following form:
\begin{equation}\label{eq31}
\begin{aligned}
&\Big(1+2\lambda H^2\Big)\ddot{A}+\Big(3H+4\lambda H \dot{H}+6\lambda H^3\Big)\dot{A}+\\
&\Big(2H^2+4\lambda H^4+\dot{H}+6\lambda \dot{H}H^2+m^2\Big)A=0.
\end{aligned}
\end{equation}
Eqs.~(\ref{eq29}), (\ref{eq30}) and (\ref{eq31}) are the background equations which we will use in the next calculations.

\subsection{Quasi de Sitter expansion}
In order to obtain an inflationary solution with this model, first, we explore the case for a quasi de Sitter expansion, namely,
\begin{equation}\label{eq32}
\dot{H}\approx 0.
\end{equation}
If we set $\dot{H}=0$ in the background equations (\ref{eq29}), (\ref{eq30}) and (\ref{eq31}), is easy  to show that
\begin{equation}\label{eq33}
\dot{A}\approx 0,
\end{equation}
and using these last two approximations, Eqs.~(\ref{eq32}) and (\ref{eq33}), it is straightforward to deduce that the vector field satisfy an algebraic equation involving $H$ and $A$ only:
\begin{equation}\label{eq37}
\frac{16 m^4 A^2 H^2\left(2H^2-m^2A^2\right)}{1+6\lambda H^2}=\left[4H^2+8 \lambda H^4m^2 A^2\left(1+2\lambda H^2\right)\right]^2.
\end{equation}
Solving this equation we obtain an expression of the form $A=f(H)$. Therefore, the background equations only admit a constant $A$ as an inflationary solution (since Eq.~(\ref{eq32}) implies that  $H$ is a constant).

\subsection{Slow-roll approximation}
Usually, the inflationary universe models are based upon the possibility of slow evolution of some scalar
field $\phi$ in a potential $V(\phi)$. Although some exact solutions of this problem exist, most
detailed studies of inflation have been made using numerical integration, or by employing
an approximation scheme. The most widely  used approximation scheme in the literature is the slow-roll approximation, which neglects the most slowly changing terms in the equations of motion \cite{liddle}.\\

\noindent Before considering the slow-roll approximation for the model under study, let's rewrite  Eq.~(\ref{eq31}) in a more convenient and suggestive form, as follows:
\begin{equation}\label{eq40}
\ddot{A}+3H \dot{A}+m_{\text{eff}}^2 A=\lambda \alpha,
\end{equation}
where $\alpha$ and  $m_{\text{eff}}$ (the effective mass of the vector field)  are defined as
\begin{equation}\label{eq41}
\alpha\equiv-2 \left[ H^2 \ddot{A}+\left(2 H \dot{H} +3 H^3\right)\dot{A}\right],
\end{equation}
and 
\begin{equation}\label{eq42}
m_{\text{eff}}^2\equiv2H^2+4\lambda H^4+\dot{H}+6\lambda \dot{H}H^2+m^2.
\end{equation}
Now, we introduce for the model  the slow-roll conditions, given by
\begin{equation}\label{eq39}
\dot{A}^2\ll m^2A^2,  \ \ \  |\ddot{A}| \ll 3H|\dot{A}|, \ \ \ 2\kappa^4|HA\dot{A}|\ll 1, \ \ \ |\dot{H}|\ll H^2.
\end{equation}
Using this slow-roll conditions in Eq.~(\ref{eq41}), we obtain
\begin{equation}\label{eq43}
\alpha\simeq 0,
\end{equation}
and Eq.~(\ref{eq40}) is reduced to
\begin{equation}\label{eq44}
\ddot{A}+3H \dot{A}+m_{\text{eff}}^2 A\simeq 0,
\end{equation}
which is similar to the equation of motion for the inflaton field in theories with scalar fields minimally coupled to gravity.\\

\noindent  In a similar way, the slow-roll conditions reduce Eqs.~(\ref{eq29}) and (\ref{eq44}) to:
\begin{equation}\label{eq46}
\begin{cases}
2H^2\simeq\kappa ^2 \left(m^2+H^2+6\lambda H^4\right)A^2,\\
3H\dot{A}+\left(2H^2+4\lambda H^4+m^2\right)A \simeq 0.
\end{cases}
\end{equation}

\noindent  It is well known that inflationary period must have a duration that allows  resolving the main problems present in the Hot Big Bang model of the universe. In this sense, a common practice is to quantify the amount of inflation using a simpler e-fold number, $N$, which measures the growth of the scale factor alone. For a successful  inflationary period, the duration of this regimen should be more than 60 e-folds  before the end of inflation.\\

\begin{figure}[t]
\centerline{\includegraphics[width=1\linewidth,scale=1]{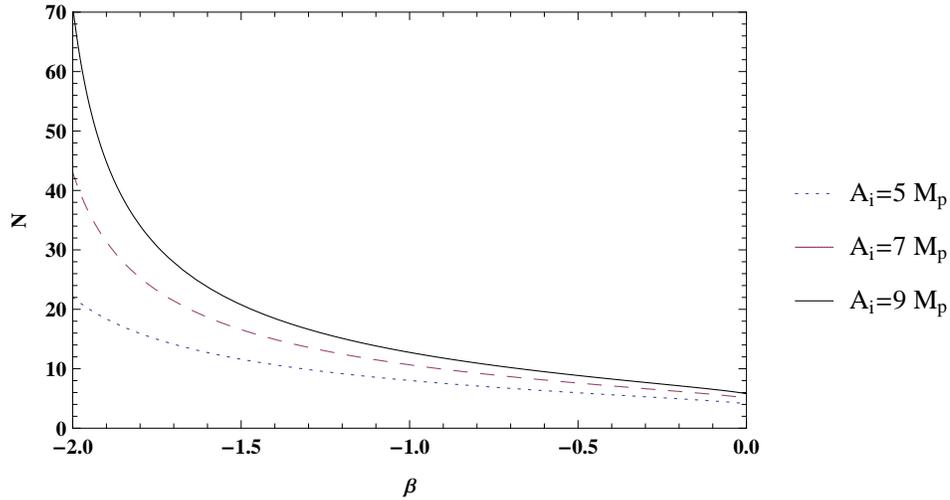}}
\caption{Evolution of the e-folding number $N$ with the parameter $\beta$, using $A_e=10^{-6}\,M_p$ and three different values for $A_i$:  $A_i=5\, M_p$ (dotted line), $A_i=7\, M_p$ (dasshed line) and  $A_i=9\, M_p$ (solid line). \label{fig1}}
\end{figure}

\noindent The e-folding number is defined by $N \equiv\int_{t_i}^{t_e}{Hdt}$, and for the inflationary model under consideration in the slow-roll approximation, using  Eqs.~(\ref{eq46}), is given by \footnote{The analytical expression for the e-folding number $N$ is not shown, since it is very large.}
\begin{equation}\label{eq47}
N =\int_{A_i}^{A_e}{\frac{H}{\dot{A}}dA}=\int_{\widetilde{A} _i^2}^{\widetilde{A} _e^2}{\frac{-9}{10+7\widetilde{A}^2-
\sqrt{(\widetilde{A}^2-2)^2-24(\widetilde{A}^2)^2\beta}}d(\widetilde{A}^2)},
\end{equation}
where  $\beta\equiv m^2\lambda $  is a dimensionless  parameter and the field  $\widetilde{A}$  is defined as $A\equiv\widetilde{A}M_p$. 
The condition that $N$ must be a real number, implies that $-2<\beta\leq 0$ (for $\beta=-2$, $N$ is indefinite). In Fig.~\ref{fig1}, we have plotted  $N$ vs. $\beta$, using $A_e=10^{-6}\,M_p$ and three different values for $A_i$ at the Planckian scale (namely, $A_i\backsim M_p$). On the other hand, making $\beta=0$ in Eq.~(\ref{eq47}) and for: (a)  $\widetilde{A}^2\geq 2$ and (b)  $0\leq\widetilde{A}^2<2$ (since,
$\sqrt{(\widetilde{A}^2-2)^2}=|\widetilde{A}^2-2|$),  we get 
\begin{equation}\label{equa48}
\text{(a)}\ \ N=\frac{3}{2}\ln{\left(\frac{\widetilde{A}_i^2+2}{\widetilde{A}_e^2+2}\right)} \ \ \text{and}\ \
\text{(b)}\ \ N=\frac{9}{8}\ln{\left(\frac{\widetilde{A}_i^2+1}{\widetilde{A}_e^2+1}\right)},
\end{equation}
and choosing the appropriate initial and ending conditions, e.g, using in (a):  $A_i=1000\,M_p$ and $A_e=2\, M_p$
($A_e\ll A_i$), one has $N\simeq 18$; and in (b), using $A_i=1\,M_p$ and $A_e=10^{-6}\, M_p$, one has $N\simeq 0.8$. We can see that in both cases it's not possible to obtain a suitable amount of e-foldings ($N\gtrsim 60$) for sufficient inflation. In addition, making $\lambda=0$ and $m=0$ (namely,
$\beta=0$) in Eqs.~(\ref{eq29}) and (\ref{eq30}) is easy to show that the present model represents radiation with the equation of state $p_A=\rho_A/3$ and in this case there's not inflation at all, which is in agreement with the above analysis.  From  Fig.~\ref{fig1} is clear that, choosing the appropriate values for $A_i$ and $\beta$, it is possible to obtain the sufficient amount of e-foldings, e.g,  $A_i=9\,M_p$, $A_e=10^{-6}\, M_p$ and $\beta=-1.97$, one has $N\simeq 60$.\\

\noindent Now, in order to verify the validity of the slow-roll approximations used previously, we have checked numerically (using the background equations (\ref{eq29}) and (\ref{eq30})) the behavior of the slow-roll parameters $\epsilon$, $\delta$, $\eta$ and $\xi$
defined as follows:
\begin{equation}\label{eq49}
\epsilon\equiv -\frac{\dot{H}}{H^2}, \ \ \ |\epsilon|\ll 1; \ \ \ \delta\equiv\frac{\dot{A}^2}{m^2A^2}, \ \ \ |\delta|\ll 1
\end{equation}
and
\begin{equation}\label{eq50}
\eta\equiv -\frac{\ddot{A}}{3H\dot{A}}, \ \ \ |\eta|\ll 1; \ \ \ \xi\equiv -2\kappa^4HA\dot{A}, \ \ \ |\xi|\ll 1.
\end{equation}

\begin{figure}[t]
\includegraphics[width=1\linewidth,scale=1]{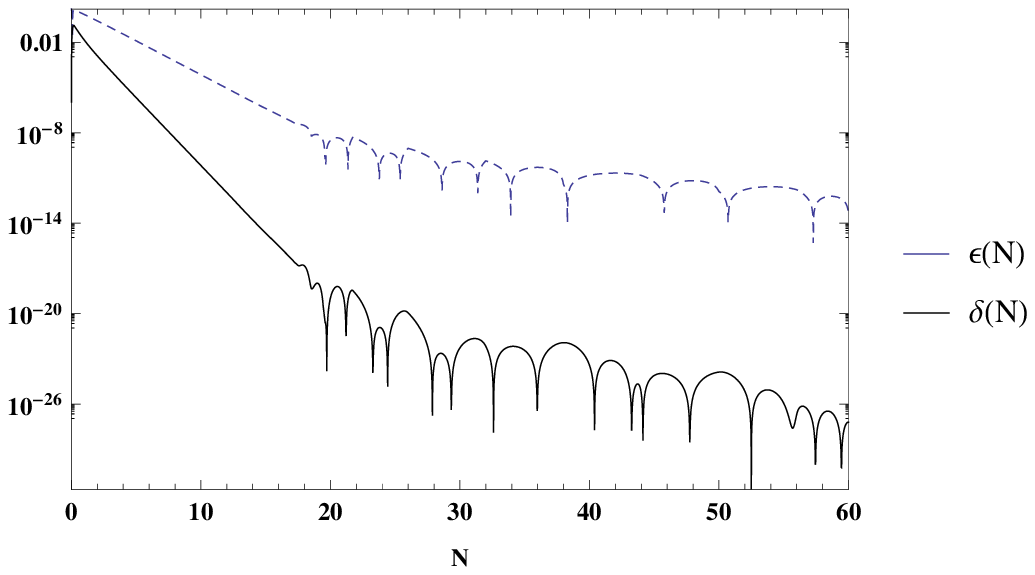}
\quad
\includegraphics[width=1\linewidth,scale=1]{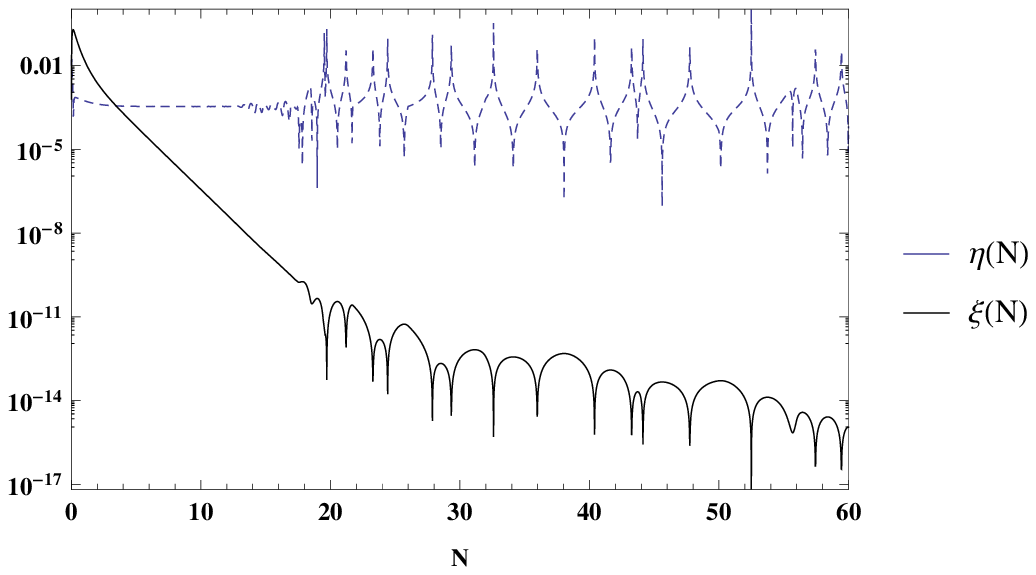}
\caption{Evolution of the slow-roll parameters  with the e-folding number $N$: (top) $\epsilon$ (dashed line) and $\delta$ (solid line);
(bottom) $\eta$ (dashed line) and $\xi$ (solid line). In this numerical simulation we have used $\lambda=-1\,M_{p}^{-2}$ and
$m^2=1\,M_{p}^{2}$ (an analogous behavior it is obtained using $\lambda=1\,M_{p}^{-2}$ and
$m^2=-1\,M_{p}^{2}$). Furthermore, the change of variable $dN=Hdt$ was performed and the initial conditions used were: $H_i=10^{-4}\,M_{p}$, $A_i=1\,M_{p}$ and  $A'_i=0$, where a prime represents a derivative with respect to e-folding variable, $N$. \label{fig2}}
\end{figure}

\noindent We can see in Fig.~\ref{fig2} that the validity of the  slow-roll parameters (namely, $|\epsilon|\ll 1$  $|\delta|\ll 1$, $|\eta|\ll 1$ and $|\xi|\ll 1$) during the inflationary phase are satisfied. \\

\noindent Finally, although the model presented here has second order field equations, this is not enough
to guarantee a stable theory. Therefore, in order to guarantee the full stability of the inflationary model considered above, it's necessary to dismiss the presence of ghosts and Laplacian instabilities. Let us recall that a ghost is a field with negative kinetic energy and a Laplacian instability implies that we have negative squared propagation speed for high enough frequencies. In general the ghosts  are associated with the longitudinal vector polarization present in  vector-tensor models (e.g, some specific models involving massive vector fields), and are found from studying the sign of the eigenvalues of the kinetic matrix for the physical perturbations. And,  the Laplacian instabilities are found from studying the sign of the squared speed of sound
(namely,  to avoid the Laplacian instability we must require that $c_s^2\geq 0$, which ensures that the equation of motion for the vector field takes the form of a wave equation and, thus, no exponentially growing solutions exist).  In this sense, it's necessary to collect all the conditions we need to impose to  the present model in order to stabilize the system around the de Sitter solution against ghosts and Laplacian instabilities. From this point of view, the strength and sign that could take the  non-minimal coupling parameter $\lambda$ is fundamental in this case, and on the other hand, the particular form for the potential used in the present model (Proca potencial) could be a source of potential instabilities. Hence, a detailed perturbative analysis around the de Sitter background, like those studied in Refs.~\citen{emami}, \citen{lavinia2} and \citen{lavinia12}, must be performed for this model. However, this analysis is beyond of the scope of the present work and  could be addressed later.

\section{Conclusions}\label{sec_concs}
\noindent In this work we have studied a  vector-tensor model of inflation with  massive vector fields and derivative self-interactions (inspired in the Generalized Proca Theory). The action under consideration contains a usual  Maxwell-like kinetic term, a general potential term and a term  with non-minimal derivative coupling between the vector field and gravity, via the dual Riemann tensor. In this theory, the last term contains  a free parameter, $\lambda$, which quantify the non-minimal derivative coupling (see Eq.~(\ref{eq8})). In this scenario, taking into account a spatially flat FRW  universe and a general  vector field, the general expressions for the  total energy momentum tensor  and the equation of motion were obtained
(see Eqs.~(\ref{eq18}), (\ref{eq19}) and (\ref{eq20})). In this model, the isotropy of expansion was guaranteed by considering a triplet of orthogonal vector fields (see Eq.~(\ref{eq21})). Using a  Proca-type potential, a suitable inflationary regimen driven by massive vector fields was studied.
In order to obtain an inflationary solution with this model, first,  the case for a quasi de Sitter expansion was considered. In this case the vector field  behaves as a constant (see Eq.~(\ref{eq37})). For the second case,  slow-roll analysis was performed and slow-roll conditions were defined for this model, which, for suitable constraints of the model parameters ($-2<\beta\leq 0$) and for $A_i$  at the Planckian scale, it was possible to obtain the required number of e-folds ($N\simeq 60$) for sufficient inflation (see Eq.~(\ref{eq47}) and Fig.~\ref{fig1}). Also, for $\beta=0$, it was shown that the present model represents radiation (at the Planckian scale) and there's not inflation at all (see  Eq. ~(\ref{equa48})). Finally, In order to verify the validity of the slow-roll approximations, we have checked numerically  the behavior of the slow-roll parameters $\epsilon$, $\delta$, $\eta$ and $\xi$ (see  Eqs. ~(\ref{eq49}), (\ref{eq50}) and Fig.~\ref{fig2}).

\section{Acknowledgments}\label{sec_acknow}
\noindent We thank the anonymous referee for his/her quite useful comments and suggestions, which helped us improve this work. This work was supported by Universidad del Atl\'antico under project CB09-FGI2016.

\end{document}